\documentclass[12pt,letterpaper,oneside]{article}
\usepackage{amssymb,tabularx,multirow,amsmath,natbib, epsfig, threeparttable, amstext, subfigure,xcolor}
\usepackage{pdflscape}
\usepackage{afterpage}
\usepackage{capt-of}
\usepackage{rotating}

\newcommand{\blind}{1}

\newcommand{\rf}{\vskip .1in\par\sloppy\hangindent=1pc\hangafter=1
                 \noindent}

\usepackage[dvips,top=1.2in,bottom=0.65in,left=1.15in,right=1.15in,includefoot]{geometry}

\usepackage[margin=20pt,labelfont={bf,small},textfont={it,small},indention=0.5cm]{caption}

\begin{document}

\def\spacingset#1{\renewcommand{\baselinestretch}%
{#1}\small\normalsize} \spacingset{1}


\if1\blind
{
  \title{\bf Animal Movement Models for Migratory Individuals and Groups}
  \author{Mevin B. Hooten\thanks{
    \tiny Corresponding author.  Email: mevin.hooten@colostate.edu; This version accepted for publication on March 28, 2018:  Hooten, M.B., H.R. Scharf, T.J. Hefley, A. Pearse, and M. Weegman. (In Press). Animal movement models for migratory individuals and groups. Methods in Ecology and Evolution.}\hspace{.2cm}\\
    U.S. Geological Survey \\ Colorado Cooperative Fish and Wildlife Research Unit \\ Department of Fish, Wildlife, and Conservation and \\ Department of Statistics \\ Colorado State University\\
    and \\
     Henry R. Scharf\\ Department of Statistics \\ Colorado State University \\
    and \\
     Trevor J. Hefley \\ Department of Statistics \\ Kansas State University \\
    and \\
     Aaron T. Pearse \\ U.S. Geological Survey \\ Northern Prairie Wildlife Research Center \\
    and \\
     Mitch D. Weegman \\ School of Natural Resources \\ University of Missouri}
  \maketitle
} \fi

\if0\blind
{
  \bigskip
  \bigskip
  \bigskip
  \begin{center}
    {\LARGE\bf Convolutions in Animal Movement Models}
\end{center}
  \medskip
} \fi

\bigskip
\pagebreak
\begin{abstract}

\noindent  1. Animals often exhibit changes in their behavior during migration.  Telemetry data provide a way to observe geographic position of animals over time, but not necessarily changes in the dynamics of the movement process.  Continuous-time models allow for statistical predictions of the trajectory in the presence of measurement error and during periods when the telemetry device did not record the animal's position.  However, continuous-time models capable of mimicking realistic trajectories with sufficient detail are computationally challenging to fit to large data sets.  Furthermore, basic continuous-time model specifications (e.g., Brownian motion) lack realism in their ability to capture nonstationary dynamics.

\noindent  2. We present a unified class of animal movement models that are computationally efficient and provide a suite of approaches for accommodating nonstationarity in continuous trajectories due to migration and interactions among individuals.  Our approach uses process convolutions to allow for flexibility in the movement process while facilitating implementation and incorporating location uncertainty.  We show how to nest convolution models to incorporate interactions among migrating individuals to account for nonstationarity and provide inference about dynamic migratory networks.

\noindent  3. We demonstrate these approaches in two case studies involving migratory birds.  Specifically, we used process convolution models with temporal deformation to account for heterogeneity in individual greater white-fronted goose migrations in Europe and Iceland and we used nested process convolutions to model dynamic migratory networks in sandhill cranes in North America.

\noindent  4. The approach we present accounts for various forms of temporal heterogeneity in animal movement and is not limited to migratory applications.  Furthermore, our models rely on well-established principles for modeling dependent data and leverage modern approaches for modeling dynamic networks to help explain animal movement and social interaction.    

\end{abstract}

\noindent%
\vfill

\newpage

\spacingset{1.45} 

\section{Introduction}
Rapid improvement in technology has led to high-quality animal tracking (i.e., telemetry; see Appendix S1 for a glossary of terms) data that are accumulating at an incredible rate (Cagnacci et al., 2010; Kays et al., 2015).  There are not only more data being collected in more studies, but the variety of data is also increasing.  Variation exists in telemetry devices, fix rates and regularity, accuracy, types of measurement error, duration, and taxa studied.  Behavioral variation also exists within individual and taxa.  Many approaches have been developed to characterize the variation within individual animal trajectories (Hooten and Johnson, 2017b; Hooten et al., 2017).  These approaches include the use of spatial and temporal covariates and clustering methods to understand the portions of animal trajectories that indicate distinctly different patterns (e.g., Whoriskey et al., 2017).  For example, potential function specifications in stochastic differential equations (SDEs; Brillinger, 2010) have facilitated the explicit inclusion of covariates in continuous-time models.  While some discrete-time models incorporate covariates as well, they also often focus on phenomenological clustering of movement processes to infer behavioral changes over time (e.g., Morales et al., 2004; Langrock et al., 2012; McClintock et al., 2012; McKellar et al., 2014).  

The use of SDEs to infer relationships between animal movement and habitat based on telemetry data is increasing (e.g., Gurarie et al., 2017b; Parton and Blackwell, 2017; Russell et al., 2017), but the associated computational challenges are also increasing as a function of data set size as well as data and model complexity (Scharf et al., 2017).  Thus, several approaches have been developed to utilize predicted continuous-time trajectories based on telemetry data in a two-stage modeling framework to infer relationships between animal movement and habitat (Hooten et al., 2010; Hanks et al., 2015; McClintock et al. 2017; Scharf et al., 2017).  In this framework, the first stage predicts the trajectory of the individual and the second stage uses that prediction to obtain inference for the effects of covariates on movement while accounting for uncertainty in the predicted trajectory in continuous-time (Hooten et al., 2017).  Thus, accurate continuous-time models are essential to represent the predicted trajectory distribution in the first stage of such approaches.  

Accurate representations of the predicted trajectory distribution may be obtained from movement models that are continuous and allow for variation in the movement dynamics throughout the trajectory.  Thus, movement models should allow for nonstationarity, a term used in time series and spatial statistics, to account for changes in the dynamics of the animal as it moves.  Nonstationarity could be caused by behavioral responses to the environment, diurnal cycles, or interactions with other individuals of the same or different species (Auger-M\'eth\'e et al., 2016; Soleymani et al., 2017).  For migratory animals specifically, heterogeneity is a natural component of their life history and leads to nonstationary dynamics in their movement trajectories (Cagnacci et al., 2016; Gurarie et al., 2017a).  Several approaches have been used to characterize migrations (Bauer and Klaassen, 2013), in both terrestrial (e.g., Fleming et al., 2014) and aquatic (e.g., Hays et al., 2016) systems.  However, key questions about animal ecology remain that new methods for analyzing telemetry data must be developed to answer (Hays et al., 2016).    

In what follows, we demonstrate a unified framework to account for nonstationarity in animal trajectories using statistical models that are flexible and computationally efficient to implement.  We show that single- and multiple-individual continuous-time models can accommodate heterogeneity in movement due to migratory processes.  We use process convolutions (Higdon, 2002) to specify flexible movement models mechanistically and nest them in a hierarchical statistical framework to properly account for measurement error.  Process convolutions have become popular in spatial statistics because they result in models that are easy to specify and fit to data (e.g., Barry and Ver Hoef, 1996).   

We illustrate our approach to characterizing nonstationary animal trajectories through two examples involving migratory birds.  First, we apply individual-based process convolution models to account for heterogeneity in the migration trajectories of Greenland white-fronted geese (\emph{Anser albifrons flavirostris}; a subspecies of greater white-fronted goose; Dalgety and Scott, 1948) from Ireland to staging grounds in Iceland.  For this example, we developed a temporal deformation for migratory animals that provides inference about the timing and duration of migration-induced nonstationarity in the movement dynamics.  In the second example, we demonstrate a nested process convolution approach that utilizes simultaneous telemetry data from multiple sandhill crane (\emph{Antigone canadensis}) individuals migrating across North America.  Our convolution-based approach provides substantial reductions in uncertainty for trajectory estimates by borrowing strength across individuals using a dynamic network specification (Jacoby and Freeman, 2016).      

\section{General Statistical Framework}
\begin{figure}[p]
  \centering
  \includegraphics[width=6in, angle=0]{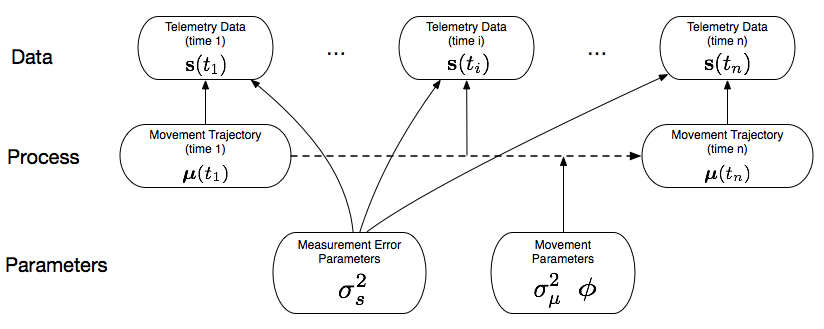}
  \caption{A directed acyclic graph (DAG) depicting the hierarchical model for telemetry data $\mathbf{s}(t_i)$ for $i=1,\ldots,n$, containing an underlying trajectory $\boldsymbol\mu(t)$ (with dashed arrow indicating the continuous-time process) and both data and process parameter sets $\sigma^2_s$, $\sigma^2_\mu$, and $\phi$. This DAG represents a basic hierarchical model for the single-individual case.}
  \label{fig:dag}
\end{figure}

Our approach can be summarized as a hierarchical model for telemetry data $\mathbf{s}(t_i)$ (a $2\times 1$ vector), for $i=1,\ldots,n$ (Figure~\ref{fig:dag}).  By constructing the statistical movement model hierarchically, we consider the mechanisms that give rise to data and underlying movement process $\boldsymbol\mu(t)$ conditionally.  After constructing the hierarchical model, we fit the model to data using an efficient computer algorithm based on an integrated likelihood (where $\boldsymbol\mu(t)$ is integrated out) that takes the form of a Gaussian process model with measurement error and temporal dependence in the trajectory accommodated by covariance.  We predict the correct latent process $\boldsymbol\mu(t)$ using a separate procedure that relies on output from the model fit.

The hierarchical Bayesian framework (Berliner, 1996; Hobbs and Hooten, 2015) we rely on contains three components (Figure~\ref{fig:dag}):  a data model (i.e., measurement error model), a process model (i.e., the movement model), and a parameter model (i.e., priors expressing our existing knowledge about model parameters).  The second stage (i.e., process model) of our hierarchical framework characterizes the movement trajectory using process convolutions.  Process convolution specifications allow us to build movement models mechanistically, based on first-order mean structure, but fit them to data using second-order covariance structure for computational efficiency (Hefley et al., 2017).  Therefore, we describe process convolutions first and then describe how to arrive at custom covariance functions using process convolutions.   

\subsection{Process Convolution Models}
Process convolutions can be thought of as moving averages (Barry and Ver Hoef, 1996; Higdon, 2002; Peterson and Ver Hoef, 2010).  If we average over a continuous-time stochastic process like white noise in a certain way, the result is an appropriately smooth (and maybe heterogeneous) trajectory that can serve as a model for movement.  

Hooten and Johnson (2017a) proposed movement models as convolutions with white noise              
\begin{equation}
  \boldsymbol\mu(t) = \boldsymbol\mu_0+\int_{t_1}^{t_n} \mathbf{H}(t,\tau)d\mathbf{b}(\tau) \;, 
  \label{eq:simpconv2}
\end{equation}
\noindent where $\boldsymbol\mu(t)$ is a $2\times1$ vector that represents the true unobserved animal position at time $t$ ($t_1 \leq t \leq t_n$, where $t_1$ and $t_n$ bound the temporal period of interest) and $d\mathbf{b}(\tau)$ is a two-dimensional scaled white noise process (scaled by $\sigma^2_\mu$).  The matrix $\mathbf{H}(t,\tau)$ in Eqn~\ref{eq:simpconv2} is a $2\times 2$ diagonal matrix with diagonal elements equal to $h(t,\tau)=\int_{\tau}^{t_n} g(t,\tilde\tau) d\tilde\tau$ (also called a ``basis function;'' e.g., Hefley et al. 2017). The function $g(t,\tau)$ is a one-dimensional temporal kernel anchored at time $t$ (e.g., a Gaussian function with location $t$).  Different choices for $g(t,\tau)$ represent different hypotheses about the ecology of the species under study (see Fig. 2 in Hooten and Johnson, 2017a).  For example, both Brownian motion and integrated Brownian motion (i.e., correlated random walk models; Johnson et al., 2008; Gurarie and Ovaskainen, 2011) can be expressed as Eqn~\ref{eq:simpconv2}.  

\begin{figure}[p]
  \centering
  \includegraphics[width=6in, angle=0]{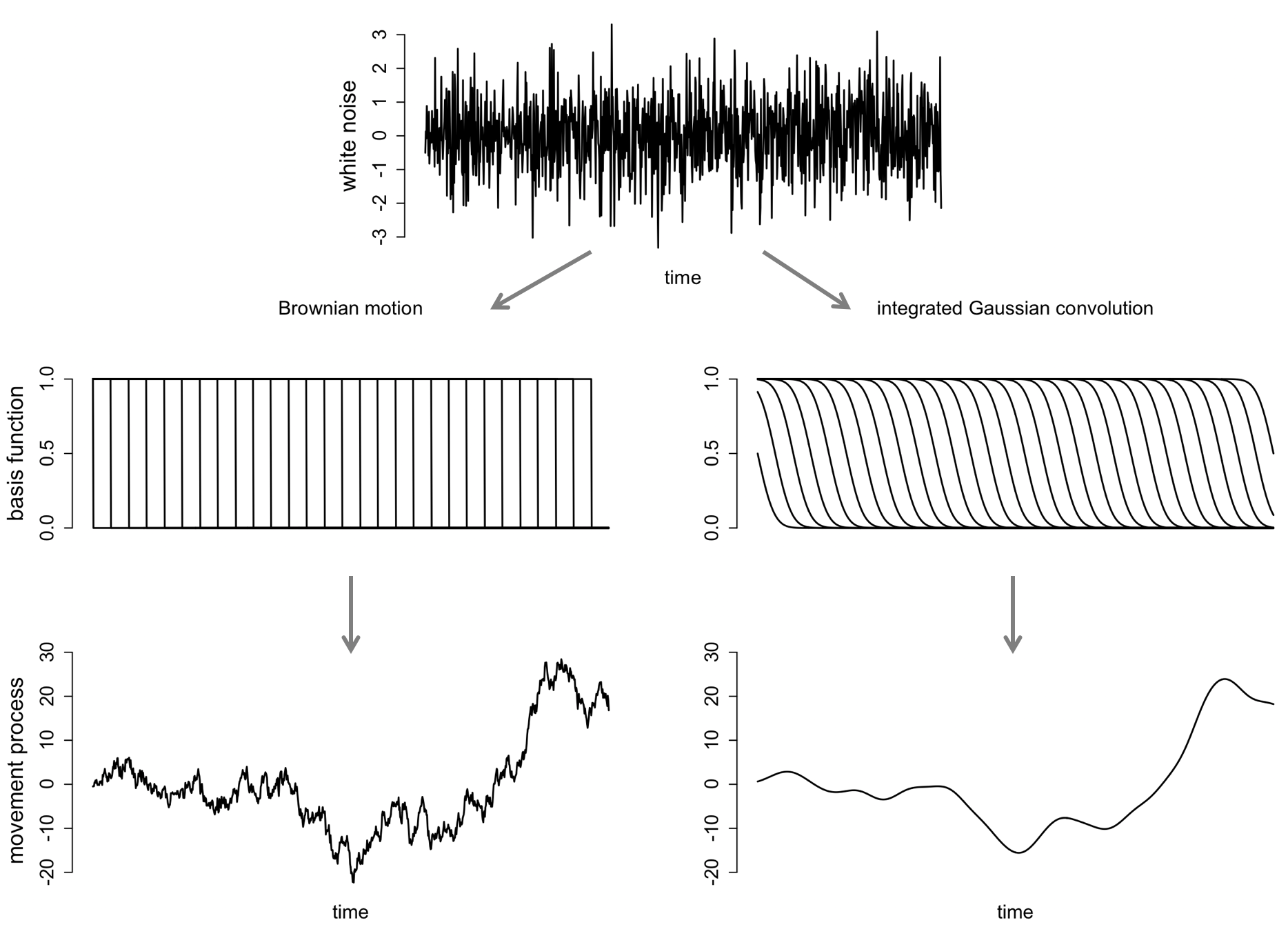}
  \caption{A one-dimensional example illustrating how white noise (top) can be convolved with Brownian basis functions (left middle) or integrated Gaussian basis functions (right middle) to yield a Brownian motion (left bottom) or integrated Gaussian movement process (right bottom).  Note that the change in scale from white noise to the movement process of interest is controlled by $\sigma^2_\mu$.}
  \label{fig:BM_fig}
\end{figure}

Eqn~\ref{eq:simpconv2} is referred to as a process convolution because it defines the position of an individual at time $t$ as a convolution (i.e., an integral of a product) of a kernel function with a stochastic process (Higdon, 2002).  In the animal movement context, process convolutions induce a form of inertial smoothing in the trajectory.  Figure~\ref{fig:BM_fig} depicts how trajectories arise (in one dimension) based on different choices of $g(t,\tau)$.  
The middle row of Figure~\ref{fig:BM_fig} represents two possible basis functions $h(t,\tau)$ at a subset of time points.  To obtain the movement process (i.e., the locations of the individual) in the bottom row of Figure~\ref{fig:BM_fig}, we multiply $h(t,\tau)$ by white noise and integrate over the time domain.  

While the exact convolution can be written in continuous-time (Eqn~\ref{eq:simpconv2}), we approximate it using numerical integration as a sum of the product, i.e., $\sum_{\cal T} \mathbf{H}(t,\tau)d\mathbf{b}(\tau)$.  The set of times ${\cal T}$ over which the sum is calculated is chosen to be large enough to provide an accurate approximation to the convolution, but small enough to still be computationally tractable.  This is the same type of approximation used in differential equation models (e.g., Cangelosi and Hooten, 2009) and integral projection models (e.g., Easterling et al., 2000; Ellner and Rees, 2006). 

Convolutions of white noise are well-studied, have useful properties (Barry and Ver Hoef, 1996; Higdon, 2002), and are attractive because they allow the user to model the system with a dynamic forward process that aligns with their hypotheses of animal movement mechanisms.  Also, convolutions can be used to construct complex covariance functions for dependent processes that are not easy to specify directly (Ver Hoef and Peterson, 2010).  In modern spatial statistics, second-order covariances are commonly parameterized using first-order representations of the dependence structure in the form of convolutions (Sampson, 2010; Hefley et al., 2017) because of the added flexibility and computational efficiency in many cases.  Our approach to movement modeling allows the user to construct models based on mechanisms, but then fit the models to data using Gaussian processes with dependence expressed via covariance. 

\subsection{Convolution-Induced Covariance}
Convolutions such as Eqn~\ref{eq:simpconv2} provide an intuitive way to specify dependence for continuous processes based on covariance (Hefley et al., 2017).  For example, convolutions provide a formal way to accommodate correlated random walk models that have a long history of use in studies of animal movement.  In spatial statistics, it is common to express dependence in terms of covariance, at least in part, because it can yield computational advantages for fitting models to data (Hefley et al., 2017).  For the convolution model in Eqn~\ref{eq:simpconv2}, we can write the covariance between time points $t_i$ and $t_j$ as $\text{cov}(\boldsymbol\mu(t_i),\boldsymbol\mu(t_j))=\int_{t_1}^{t_n} \sigma_\mu^2 \mathbf{H}(t_i,\tau)\mathbf{H}(t_j,\tau)'d\tau$ (where $'$ denotes a transpose).  
This allows us to construct the full hierarchical model (Figure~\ref{fig:dag}) with data modeled conditionally as $\mathbf{s}(t_i) \sim \text{N}(\boldsymbol\mu(t_i),\sigma^2_s \mathbf{I})$, where $\sigma^2_s$ represents telemetry error variance, and the continuous-time trajectory $\boldsymbol\mu$ conditionally modeled as a Gaussian process with mean $\boldsymbol\mu_0$ and covariance as specified above.  

To fit the model, we integrate out $\boldsymbol\mu$ to yield a Gaussian process model for the observed telemetry data $\mathbf{s}\equiv (s_1(t_1),\ldots,s_1(t_n),s_2(t_1),\ldots,s_2(t_n))'$ directly as
\begin{equation}
  \mathbf{s} \sim \text{N}(\boldsymbol\mu_0\otimes \mathbf{1}, \sigma^2_s\mathbf{I} + \sigma^2_\mu (\mathbf{I}\otimes \mathbf{H})(\mathbf{I}\otimes \mathbf{H})')\;, 
  \label{eq:jointmod}
\end{equation}
\noindent where, the $n\times m$ matrix $\mathbf{H}$ has $(i,j)^{\text{th}}$ element $h(t_i,\tau_j)$ computed at a subset of $m$ times associated with the finite approximation of the integral as previously described.  The integrated model in Eqn~\ref{eq:jointmod} now accommodates the data and process levels of the hierarchical model in Figure~\ref{fig:dag} simultaneously.  The multivariate Gaussian form of the jointly specified movement model in Eqn~\ref{eq:jointmod} is attractive because efficient numerical methods can be used to fit the model to high-resolution telemetry data sets.  Many approaches for fitting Gaussian process models efficiently have been developed for use in spatial statistics, including reduced rank methods (Wikle, 2010), predictive processes (Banerjee et al., 2008), covariance tapering (Furrer, 2006), nearest neighbor methods (Datta et al., 2016), among others.     

\subsection{Temporal Deformation}
The statistical model in Eqn~\ref{eq:jointmod} is flexible because the shape, range, and scale of the kernels $g(t,\tau)$ can vary with time $t$ to accommodate realistic dynamics in animal movement.  For example, in a Gaussian kernel, the temporal range parameter ($\phi$) may vary (i.e., $g(t,\tau)\propto \exp(-(t-\tau)^2 / \phi(t))$) to allow for heterogeneity in the smoothness of the individual's track over time (Higdon, 2002).  Larger values for $\phi(t)$ imply animal behavior that results in smoother trajectories, such as migration periods, and as $\phi(t)$ decreases toward zero, the process becomes less smooth.  An alternative to letting the range parameter vary in the kernel function, is to deform (i.e., compress or expand) the temporal domain itself (Sampson and Guttorp, 1992).  By compressing time in certain regions and expanding it in others, we can account for the same type of heterogeneous dynamics as the varying parameter approach previously described.  By conditioning on a temporal deformation, we are able to use the same software we would use to fit the temporally homogeneous convolution model.      

Temporal deformation can be induced using a warping function $w(t)$ in place of time (e.g., $g(t,\tau)\propto \exp(-(w(t)-\tau)^2 / \phi)$).  Warping functions have traditionally been expressed as smooth stochastic functions in the time domain, such as Gaussian processes (e.g., Hooten and Johnson, 2017a).  The derivative of the warping function (i.e., $dw(t)/dt$) indicates the portions of the time domain that are compressed ($dw(t)/dt < 1$) and expanded ($dw(t)/dt > 1$).  Temporal compression leads to rough trajectories and temporal expansion leads to smooth trajectories.  Gaussian process warping functions are quite general, but not mechanistically linked to known natural history or animal behavior.  Furthermore, a critical characteristic of deformation approaches to account for nonstationarity is that the warping function does not fold (i.e., the resulting warped time field retains the same order as the original time domain).  Previous implementations of deformation approaches have imposed a non-folding constraint by tuning the Gaussian process associated with the warping function so that it does not result in temporal expansions that induce folding.  Such constraints may be computationally demanding to implement using conventional deformation approaches.  

\begin{figure}[p]
  \centering
  \includegraphics[width=6in, angle=0]{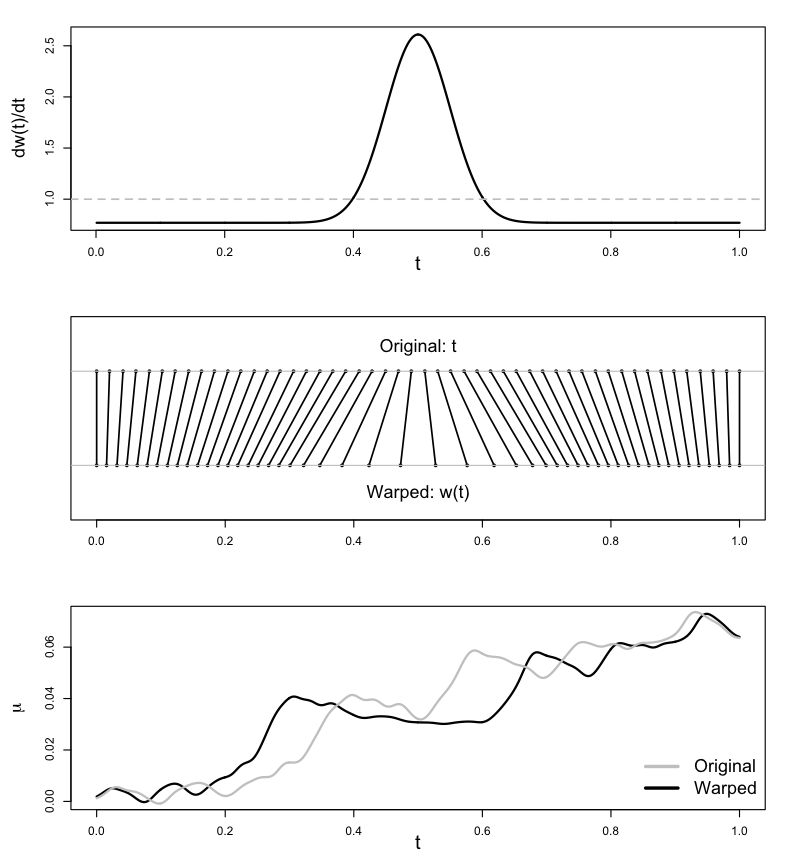}
  \caption{Top panel: An example warp function derivative ($dw(t)/dt$) based on a truncated Gaussian function f(t) anchored at $t=0.5$.  The dashed horizontal gray line indicates where the warp derivative equals 1 (indicating the inflection points delineating temporal compression or expansion) when the horizontal line intersects $dw(t)/dt$.  Middle panel:  The mapping from the original time ($t$) to the warped time ($w(t)$). Bottom panel:  Simulated one-dimensional trajectories ($\boldsymbol\mu$) based on the original time (gray) and warped time (black).}
  \label{fig:warp_fig}
\end{figure}

We propose a warping function for implementing temporal deformation that acknowledges the natural history of migratory animals and is guaranteed not to fold.  We refer to this warping function as a ``temporally deforming cumulative function'' (TDCF).  The TDCF stretches time appropriately to provide inertial smoothness in the trajectory during migration and is defined as 
\begin{equation}              
  w(t) = \frac{\sigma^2_w F(t)+ t - t_1}{\sigma^2_w + t_n - t_1} \;,
  \label{eq:tdcf}
\end{equation}              
\noindent where $\sigma^2_w\geq 0$, $F(t)=\int_{t_1}^t f(\tau) d\tau$, $f(t)$ is any non-negative function (e.g., a probability density function) that integrates to one over the time domain, and $t_1$ and $t_n$ represent the beginning and end of the time domain (or first and last times at which data were collected).  It can be shown that the derivative of the TDCF in Eqn~\ref{eq:tdcf} is 
\begin{equation}              
  \frac{dw(t)}{dt} = \frac{\sigma^2_w f(t) + 1}{\sigma^2_w + t_n - t_1}\;,
  \label{eq:tdcf_deriv}
\end{equation}              
\noindent which is a linear function of $f(t)$. 
The general form of TDCF in Eqn~\ref{eq:tdcf} will not fold the time domain and will retain the original temporal extent of the data (Figure~\ref{fig:warp_fig}).  The latter characteristic can be helpful for specifying a prior distribution for the range parameter ($\phi$) in Bayesian implementations of the convolution model in Eqn~\ref{eq:jointmod}.  

Finally, we note that the temporal deformation approach could be employed in most continuous-time models.  For example, the correlated random walk model proposed by Johnson et al. (2008) is a member of the class of models we describe herein, thus the same approach to account for temporal heterogeneity using a TDCF applies there as well.  We return to specific forms of deformation and warping kernel functions in an example involving Greenland white-fronted geese.       

\subsection{Dynamic Movement Networks}
Temporal heterogeneity in animal movement dynamics may also arise as a result of intraspecific interactions.  While adding complexity to the statistical model, accounting for dependence among individuals as populations redistribute over space can be beneficial for inference (as we demonstrate in the sandhill crane example that follows).  Many studies collect telemetry data for multiple individuals of a population or community simultaneously. Thus, we can make use of those data to improve our understanding of the trajectories of each individual using an extension of the convolution approach.  Scharf et al. (2018) proposed a nested structure for multiple convolutions that we use to reconcile the individual-level movement model in Eqn~\ref{eq:simpconv2} with dynamic social network models for movement (e.g., Russell et al., 2016; Scharf et al., 2016; Russell et al., 2017).

The process convolution in Eqn~\ref{eq:simpconv2} is actually a two-level nested convolution with the first level resulting in Brownian motion and the second performing the inertial smoothing (Hooten and Johnson, 2017).  However, that nested convolution can be expressed as a single process convolution with white noise to obtain the covariance for the Gaussian process model in Eqn~\ref{eq:jointmod}.  Extending this concept one step further, a general three-level nested process convolution structure can be expressed as

\begin{align}
  \boldsymbol\mu_{j}^{(1)}(t) &= \sum_{k=1}^J \int_{t_1}^{t_n} \mathbf{H}_{jk}^{(1)}(t,\tau)d\mathbf{b}_k(\tau) \;, \label{eq:mu1} \\ 
  \boldsymbol\mu_{j}^{(2)}(t) &= \boldsymbol{\mu}_{0,j}+\sum_{k=1}^J \int_{t_1}^{t_n} \mathbf{H}_{jk}^{(2)}(t,\tau)\boldsymbol\mu_k^{(1)}(\tau)d\tau \;, \label{eq:mu2} \\ 
  \boldsymbol\mu_{j}^{(3)}(t) &= \sum_{k=1}^J \int_{t_1}^{t_n} \mathbf{H}_{jk}^{(3)}(t,\tau)\boldsymbol\mu_k^{(2)}(\tau)d\tau \;, \label{eq:mu3}
\end{align}
\noindent where $j$ and $k$ correspond to observed individuals and the kernel functions on the diagonals of each of the convolution matrices ($\mathbf{H}_{jk}^{(1)}(t,\tau)$, $\mathbf{H}_{jk}^{(2)}(t,\tau)$, $\mathbf{H}_{jk}^{(3)}(t,\tau)$) are specified as 

\begin{align}
  h_{jk}^{(1)}(t,\tau) &= 1_{\{\tau < t\}} 1_{\{j = k\}} \;, \label{eq:h1} \\ 
  h_{jk}^{(2)}(t,\tau) &= \exp\left(-\frac{(t-\tau)^2}{\phi}\right) 1_{\{j = k\}} \;, \label{eq:h2} \\ 
  h_{jk}^{(3)}(t,\tau) &= \frac{\nu_{jk}(t)}{\sum_{k=1}^J \nu_{jk}(t)} 1_{\{t = \tau\}} \;. \label{eq:h3} 
\end{align}

The first two kernel functions ($h_{jk}^{(1)}(t,\tau)$ and $h_{jk}^{(2)}(t,\tau)$, where $1_{\{\cdots\}}$ is an indicator function that is equal 1 when the subscript condition is met and zero otherwise) are the same as in Eqn~\ref{eq:simpconv2} from the previous example for individual-based movement (i.e., inducing Brownian and inertial smoothing).  However, the third kernel ($h_{jk}^{(3)}(t,\tau)$) is a function of a weighted network describing the joint dynamics of a group of moving individuals.  The network weights $\nu_{jk}(t)$ correspond to pairwise relationships among individuals that may vary over time.  

Many approaches have been proposed for modeling network weights, including exponential random graph models and latent space models (Goldenberg et al., 2010; Farine and Whitehead, 2015; Farine, In Press).  In what follows, we describe a latent space approach (Hoff et al., 2002) to model the network weights $\nu_{jk}(t)$ based on distances among a set of points in a latent Euclidean space ${\cal Z}$.  The latent points $\mathbf{z}_j(t)$ act as random effects in the model and require a prior (described in Appendix S3).  Modeling the latent points $\mathbf{z}_j(t)$, instead of the network weights $\nu_{jk}(t)$ directly, simplifies the parameterization of the network substantially and facilitates estimation.   
If the points $\mathbf{z}_j(t)$ and $\mathbf{z}_k(t)$ are close in latent space, the network weight between them is large.  For inference in static networks where the social relationship of the individuals is homogeneous over time, all $\mathbf{z}_j(t)=\mathbf{z}_j$ and the set of $\mathbf{z}_j$ (for $j=1,\ldots,J$) arise as a point process in ${\cal Z}$.  

In the case where we expect the social structure of the observed individuals to change over time, a variety of dynamic models are available for $\mathbf{z}_j(t)$.  We describe one such specific dynamic model for $\mathbf{z}_j(t)$ in the application pertaining to sandhill crane migrations (Appendix S3).   

Heuristically, by expressing group movement using process convolutions, we are able to account for complex dynamic dependencies within and among individuals as they move.  Process convolutions allow us to parameterize second-order covariance matrices for group movement using mechanistic first-order structure (Hefley et al., 2017).  In Appendix S3, we show that the nested process convolution in Eqns~(\ref{eq:mu1})--(\ref{eq:mu3}) results in a single Gaussian process model that resembles a geostatistical model commonly used in spatial statistics.  This linkage between first-order and second-order model formulations is well-known in environmental science (Higdon, 2002; Wikle, 2010), but is still fairly new in animal movement modeling, where its potential utility is high.    

\section{Applications}
\subsection{Individual Movement: Greenland White-Fronted Geese}
The Greenland white-fronted goose (GWFG) is the most morphologically distinct subspecies of the circumpolar greater white-fronted goose (\emph{A. albifrons}; Ely et al. 2005). GWFG are long-distance migrants that breed in west Greenland (Malecki et al., 2000), stage during autumn and spring in south and west Iceland, and winter at over 70 regularly used sites across Great Britain and Ireland (Ruttledge and Ogilvie, 1979).  Thus, their annual migration spans 5,000 km and includes crossing the Greenland Ice Sheet (a 1.7 million km$^2$ expanse of ice peaking at 3,000 m in elevation; Comiso and Parkinson 2004). The global population of GWFG has declined in recent years, from approximately 36,000 individuals in 1999 to 19,000 in 2016 (Fox et al., 2016), and poor productivity has been identified as the proximate demographic mechanism for population change (Weegman et al., 2017).  GWFG are listed as `Endangered' under IUCN Red List criteria and as a priority species in the Biodiversity Action Plan in the United Kingdom (U.K.), and managed under a Species Action Plan through the African-Eurasian Migratory Waterbird Agreement (Stroud et al., 2012).  GWFG have been protected from hunting since 1982 in Ireland and Scotland, 2006 in Iceland and 2009 in Greenland; a voluntary shooting ban on the birds remains in place in Wales, where they are still legal quarry, as in England. 

GWFG occupy breeding areas from May to early September and feed on tubers and exposed plant matter, mainly common cottongrass (\emph{Eriophorum angustifolium}; Madsen and Fox, 1981). They lay 4-6 eggs and incubation occurs over 25-27 days (Fox and Stroud, 1988), similar to other Arctic-nesting geese (Cooke et al., 1995).  A four-week complete wing moult occurs during late summer. Autumn migration begins in September and birds stage in Iceland until October (now into early November; Fox et al., 1999), when they migrate to wintering areas in Great Britain and Ireland. Food sources on staging and wintering areas are mainly agricultural (e.g., cereal crops or managed grassland; Fox and Stroud, 2002). Although spring migration from Great Britain and Ireland began in April in the 1970s and 1980s, in recent years, birds have departed for Iceland successively earlier and now do so in late March (Fox et al., 2014), with greater fat stores than in previous years (Fox and Walsh, 2012).  The spring staging period in Iceland has increased in duration over the same time period because, although GWFG arrive earlier, they depart within a few days of historical departure dates in early May (Fox et al., 2014). 

During late winter 2016, GWFG were caught over intensively managed grassland at Wexford Slobs, Ireland using rocket-propelled nets (Wheeler and Lewis, 1972) under permission from the British Trust for Ornithology.  We analyzed data from 4 female GWFG that were fitted with 28 g Global Positioning System (GPS) tracking devices (with internal GPS aerial; Cellular Tracking Technologies; Rio Grande, New Jersey, USA) attached to neck collars (i.e., total package weight $=$ 39 g). The GPS logger measured and recorded spatial position $\mathbf{s}_j(t_i)$ at each fix. Tags were programmed to log eight GPS fixes per day. Data were uploaded daily to an online user interface via the Global System for Mobile Communications technology. 

We used the convolution model (Eqn~\ref{eq:jointmod}) to analyze the GWFG telemetry data from each individual separately.  We used Gaussian kernel functions ($h(t,\tau) \propto \exp(-(w(t) - \tau)^2/\phi)$) with a prior for the range parameter $\phi$ specified as a discrete uniform distribution to facilitate computation (see Hooten and Johnson, 2017a, for details).  To account for heterogeneity in time, we applied the temporal deformation approach using the warping function in Eqn~\ref{eq:tdcf_deriv} based on the mixture model 
\begin{equation}
  \mathbf{s} \sim
  \begin{cases}
   [\mathbf{s} | \boldsymbol\theta_1] \text{ w.p. } p_1 \\
   \hspace{.2in} \vdots \\
   [\mathbf{s} | \boldsymbol\theta_L] \text{ w.p. } p_L \\
  \end{cases} \label{eq:mixture}
\end{equation}
  where $[\mathbf{s} | \boldsymbol\theta_l]$ refers to the integrated likelihood in Eqn~\ref{eq:jointmod} with $\boldsymbol\theta_l$ representing all of the model parameters for the $l$th potential warping function.
Specifically, we defined the $l$th warping function using a truncated Gaussian function with location $t^*_l$, scale parameter $\phi_l$, and support $(t_1,t_n)$ for $f_l(t)$ (associated with the $l$th element of Eqn~\ref{eq:mixture}).  Mixing over models with different warping functions allowed us to approximate nearly any temporal deformation indicated by the data and also facilitates the computational implementation because we can use Bayesian model averaging (BMA) across mixture components (estimating the mixture probabilities $p_l$ as posterior model probabilities).  We used the two-stage BMA approach by Barker and Link (2013), distributing each model fit across processors, and then recombining the results in a second stage algorithm to obtain the optimal model averaged trajectory.  See Appendix S2 for priors and further model implementation details (computer code available at 

\verb|https://github.com/henryrscharf/Hooten_et_al_MEE_2018|).       

\begin{figure}[p]
  \centering
  \includegraphics[width=6in, angle=0]{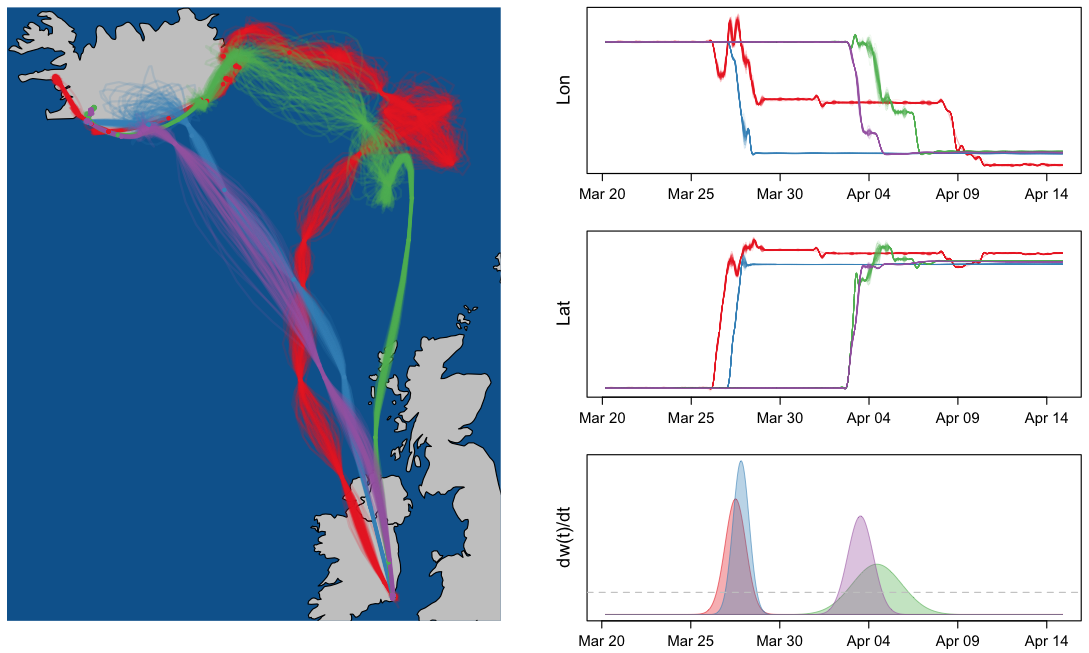}
  \caption{Left panel: Model averaged posterior predictive realizations of the geographic trajectories of 4 GWFG individuals (indicated in the colors: red, blue, green, and purple) migrating from Ireland (bottom right) to Iceland (top left).  Right panels:  Marginal trajectories corresponding to those in the left panel for longitude and latitude, respectively. Bottom right panel represents the model averaged warping function derivative $dw(t)/dt$ associated with each individual by color.  The dashed gray line illustrates when the warp derivative equals one, above which temporal expansion is indicated.}
  \label{fig:GWFG_traj}
\end{figure}

The results of fitting the model to 4 GWFG individuals are presented in Figure~\ref{fig:GWFG_traj}, where the left panel shows the estimated trajectories for the four GWFG individuals on their migration from Ireland to Iceland during March 20 - April 15, 2017.  
In Figure~\ref{fig:GWFG_traj}, the trajectories are shown as posterior predictive realizations (the individual lines) from the model averaged posterior distribution.  Figure~\ref{fig:GWFG_traj} also indicates the utility of the warp functions to account for heterogeneity in the migration of each individual.  We can also glean inference from the derivatives of the model averaged warp functions because they indicate the time and duration of the migratory period for each individual, illustrating the variability in migration among individuals.  The red and blue individuals both departed on their migration early (March 26-27) with a similar migratory duration, whereas the green and purple individuals both departed late (April 2) with the green individual taking a more circuitous route that lasted nearly twice as long.   

Convolution models provide a statistically principled means to predict true animal trajectories while accommodating uncertainty in the data and heterogeneous dynamics.  Temporal deformations that acknowledge the natural history of the species (e.g., the TDCFs we proposed herein) also provide a way to quantify differences in migration characteristics among individuals and over time.  For example, for the GWFG migration trajectories we analyzed, there appeared to be two groups in terms of timing of migration initiation, with two individuals departing in late March and the other two in early April (Figure~\ref{fig:GWFG_traj}, lower right panel).  However, there was no clear indication that the differences in migration onset were related to individual fitness or age.  In the second group of migrating individuals (purple and green individuals in Figure~\ref{fig:GWFG_traj}), the migration of the green individual led to nearly double the energetic demands as the purple individual because the total distance traveled was substantially longer (posterior mean of 3062 km for green vs.\ 1824 km for purple, during the study period).  Prior to our analyses, little evidence existed that GWFG stopover on the Faroe Islands between wintering and staging areas.  While the green individual began northward almost immediately upon departing wintering areas, the red individual seemed to lose its ability to orient correctly approximately half way through the trip (perhaps due to weather, influence of other individuals, or other unknown causes).  After a loop north of the U.K. however, the red individual corrected its orienting and reached Iceland with a total distance traveled of 4499 km and posterior mean speed of 7.28 km/hr (as compared to 1914 km [3.10 km/hr], 3062 km [4.96 km/hr], and 1824 km [2.95 km/hr] for the blue, green, and purple individuals, respectively).  Remarkably, the total migration period from wintering to staging areas for the red individual (Figure~\ref{fig:GWFG_traj}, lower right panel) was not longer than those individuals that flew directly to Iceland from Ireland (purple and blue).  However, the red individual shifted its position westward after reaching Iceland initially on the east side.       

\subsection{Group Movement: Sandhill Cranes}
Sandhill cranes (SACR; \emph{Antigone canadensis}) are a long-lived bird species found in wetland-rich landscapes across North America.  SACR are divided into various migratory and nonmigratory management populations across North America.  The midcontinent SACR population (MCP) is the largest, comprising approximately 600,000 individuals (Kruse and Dubovsky 2015).  They breed from western Quebec, across the Canadian Arctic and Alaska to northeastern Russia in a variety of ecoregions from Arctic tundra to temperate grasslands.  Twice each year SACR migrate through the Great Plains and winter from southern Oklahoma to northern Mexico, using playa and coastal wetlands for roosting and foraging (Krapu et al., 2011, 2014).   

The SACR is a species with a unique convergence of multiple user groups that share a common interest in the continued health of the species.  Midcontinent SACR are a popular sport harvest species during fall and winter in Canada and the United States and are the most hunted population of cranes in the world.  Furthermore, SACR attract a large and committed following of wildlife viewers.  For example, spring staging and courtship displays along the Platte River in central Nebraska attracts tens of thousands of people each year (Stoll et al., 2006).  

Four geographically distinct groups can be identified forming the midcontinent population.  Each expresses differences in breeding, migration, and wintering space use and timing; groups also differ in potential exposure to hunting (Krapu et al., 2011).  For this study, all individuals were from a single group that breeds in western Alaska and Siberia (lesser SACR; a subspecies distinction) and represent the smallest individuals found in the midcontinent population.  The lesser SACR group also has the greatest abundance, comprising approximately 40\% of the entire MCP (Krapu et al., 2011).

We analyzed data from 5 adult SACRs that were captured by rocket-propelled nets (Wheeler and Lewis, 1972) during March and April 2011 in the North Platte River Valley near North Platte, Nebraska.  Captured birds were tagged with a solar powered GPS platform terminal transmitter (GPS-PTT; Geotrak, Inc., Apex, North Carolina) attached with two-piece leg bands and released in the same location. GPS-PTTs can remotely provide locations to within approximately ten meters of the true position of the transmitter; therefore, they are the most accurate non-invasive tracking method available for use on this wide-ranging species.  Transmitters were programmed to record GPS locations every 6-8 hours, which provided daytime and nighttime locations, allowing for detailed information on roosting sites, diurnal use sites, and flight paths.  We monitored and archived crane locations from data provided by ARGOS (www.argos-system.org).  

We used the nested process convolution model (Eqns~\ref{eq:mu1}--\ref{eq:mu3}) described in the Dynamic Movement Networks section, which results in a Gaussian process model of the form in Eqn~\ref{eq:jointmod} (see Appendix S3 for details), to analyze the SACR telemetry data for all 5 individuals simultaneously during late summer and autumn 2013.  To account for heterogeneity in time, we specified the third kernel function (Eqn~\ref{eq:h3}) using latent space network weights $\nu_{jk}(t)=\exp(-(\mathbf{z}_j(t) - \mathbf{z}_k(t))'(\mathbf{z}_j(t) - \mathbf{z}_k(t)))$, with priors for $\mathbf{z}_j(t)$ as described in Appedix S3 (computer code available at 
\verb|https://github.com/henryrscharf/Hooten_et_al_MEE_2018|).  

\begin{figure}[p]
  \centering
  \includegraphics[width=6in, angle=0]{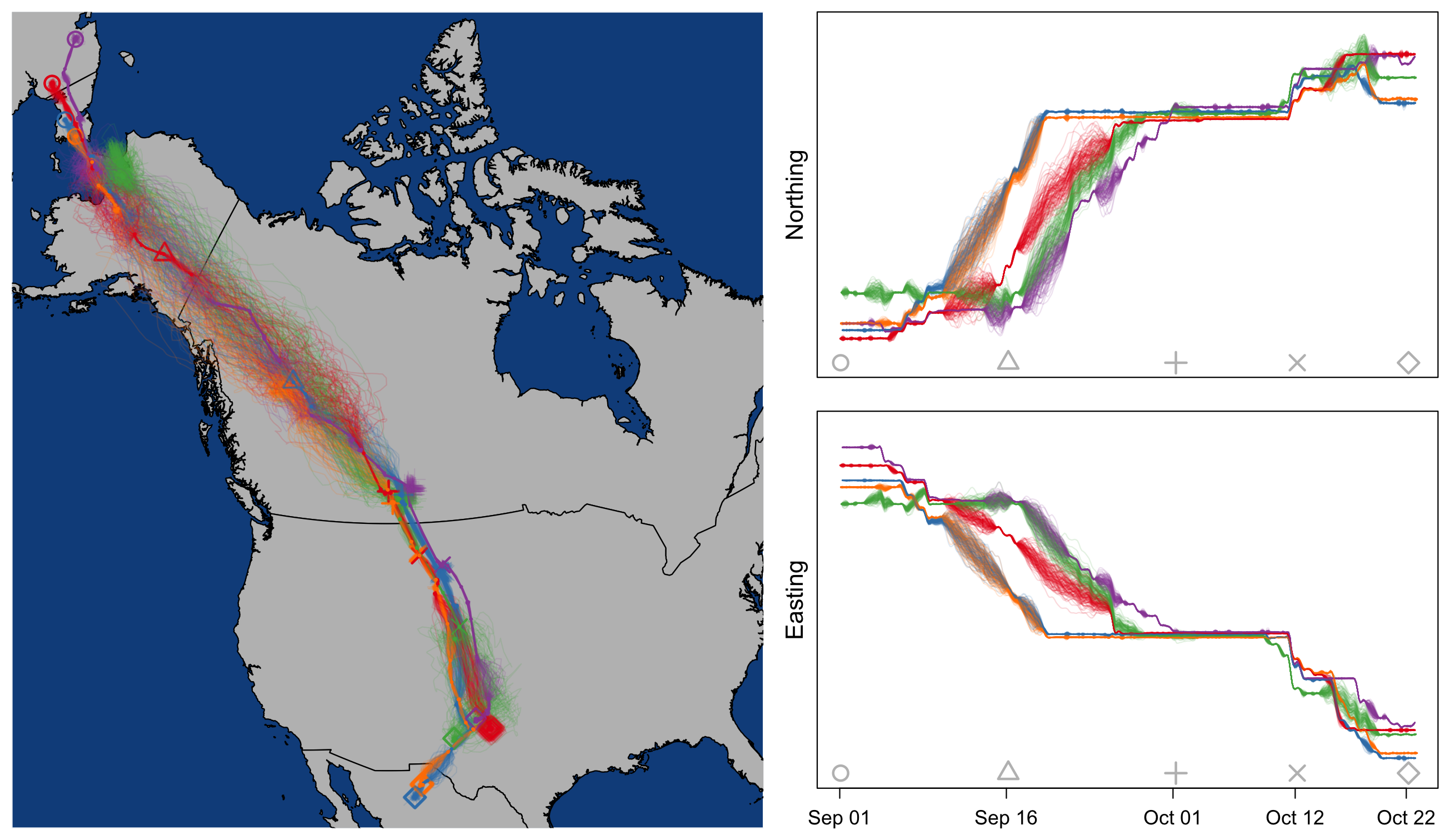}
  \caption{Left panels:  Marginal posterior predictive realizations of individual trajectories corresponding to those in the right panel for northing and easting, respectively.  Right panels: Posterior predictive realizations of the geographic trajectories of 5 SACR individuals (indicated in the colors: red, dark blue, light blue, purple, and orange) migrating from Siberia and Alaska (top left) to the southern United States and Mexico (bottom middle).}
  \label{fig:SACR_traj}
\end{figure}

The results of fitting the migratory network model to 5 SACR individuals are presented in Figure~\ref{fig:SACR_traj}, where the left panel shows the estimated trajectories for the SACR individuals in geographic space.  The right panels of Figure~\ref{fig:SACR_traj} correspond to the marginal trajectories in latitude and longitude, respectively, and the gray symbols along the x-axis are placed at the time points of the positions in geographic space on the right panel.   

\begin{figure}[p]
  \centering
  \includegraphics[width=4in, angle=0]{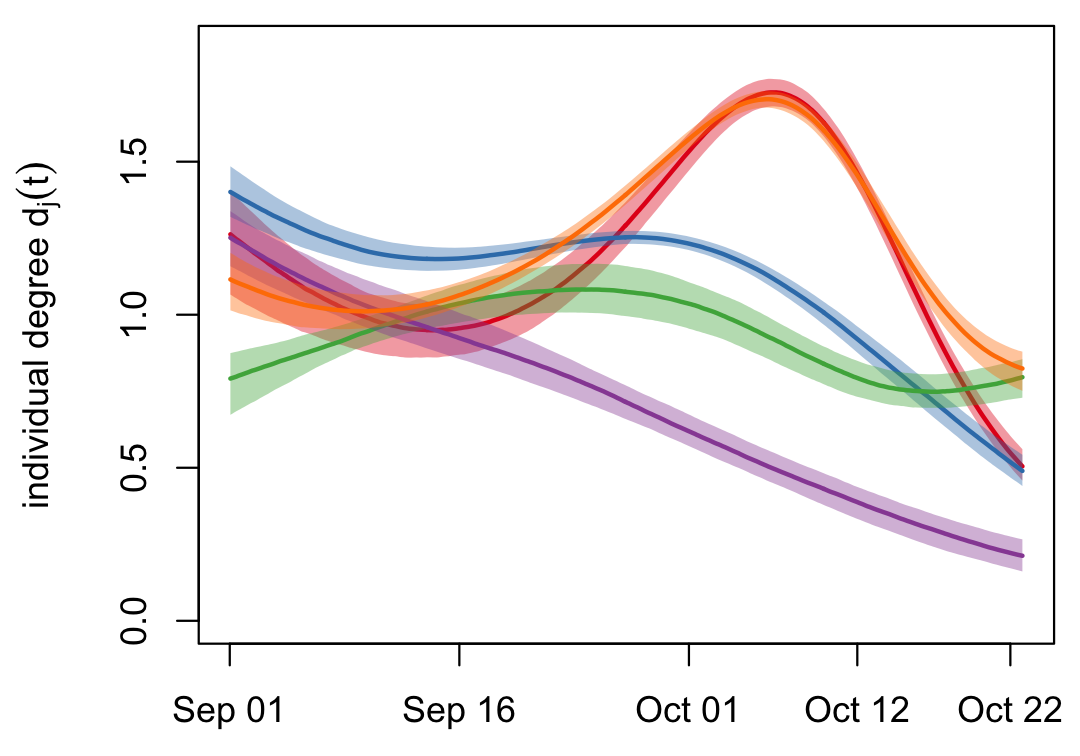}
  \caption{Individual network degree $d_j(t)$ for 5 SACR individuals during a migration from Siberia and Alaska to the southern United States and Mexico.  Large values for an individual indicate that it is more connected in the network. Colors correspond to the individual trajectories shown in Figure~\ref{fig:SACR_traj}.}
  \label{fig:SACR_deg}
\end{figure}

The trajectories in Figure~\ref{fig:SACR_traj} provide insight about the geographic positions and timing of the SACR individuals.  However, we also gain inference on the network connectivity.  Figure~\ref{fig:SACR_deg} illustrates the dynamic connectedness of each SACR individual during the migration via the derived quantity referred to as ``individual degree'' $d_j(t) = \sum_{k\neq j} \nu_{jk}(t)$.  Individual degree $d_j(t)$ is a function of the migratory network weights $\nu_{jk}(t)$, thus, we can assess its uncertainty using the Markov chain Monte Carlo (MCMC) output based on the model fit (using the equivariance property of MCMC, Hobbs and Hooten, 2015). 

In the migratory group of SACR individuals we analyzed, Figure~\ref{fig:SACR_deg} indicates that all individuals are connected to approximately one other individual in the migratory network during early September.  However, in early October, the individuals we analyzed reached the Prairie Pothole region of North America (near the $\times$ symbol in the right panel of Figure~\ref{fig:SACR_traj}).  During the week long stopover in the Prairie Pothole region, the red and orange individuals mostly stayed within a few kilometers of each other while the blue, green, and purple SACR individuals remained farther away.  SACR fly multiple kilometers daily between nocturnal roost wetlands and various diurnal foraging sites.  Thus, these daily flights imply that the red and orange individuals were likely aware of each other during this portion of the migration, but less aware of the blue, green, and purple individuals.  

\begin{figure}[p]
  \centering
  \includegraphics[width=6in, angle=0]{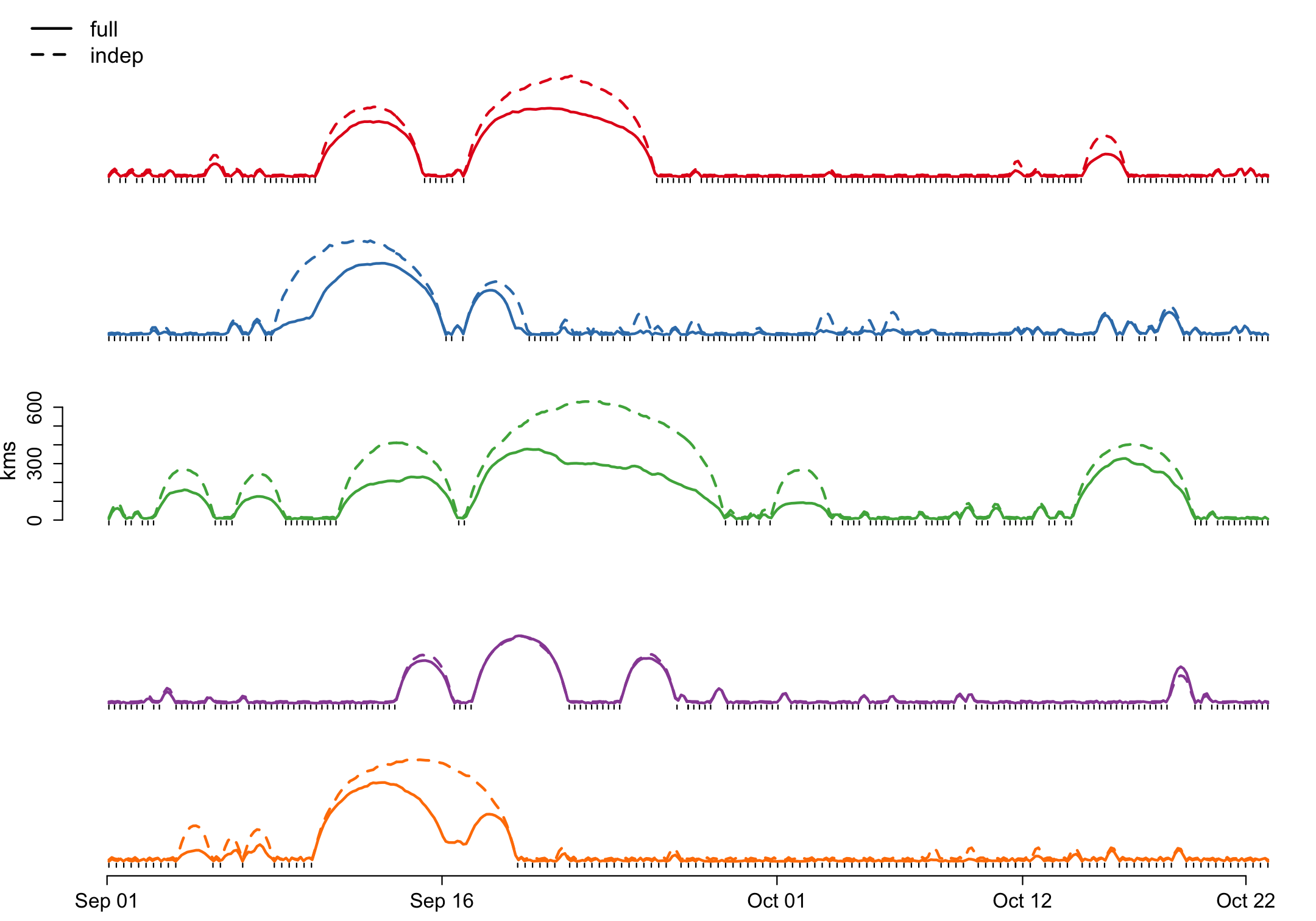}
  \caption{Location uncertainty for 5 SACR individuals over time indicated by the radius of the 95\% credible circle for $\boldsymbol\mu_j(t)$ on the y-axis.  The dashed lines represent the uncertainty inferred from models fit independently to each individual (i.e., assuming $w_{jk}(t)=0$) whereas the solid lines represent the uncertainty resulting from the migratory network model.  Colors correspond to the individual trajectories shown in Figure~\ref{fig:SACR_traj} and the dashes below each line indicate times for which data exist.}
  \label{fig:SACR_uncertainty}
\end{figure}

In addition to providing insights into the movement ecology and behavior of animals, the migratory network model we described herein can be used to reduce the uncertainty of the individual trajectories $\boldsymbol\mu_j(t)$ when individuals in a migratory group are inferred to be connected.  For example, Figure~\ref{fig:SACR_uncertainty} illustrates the reduction in uncertainty that is gained by modeling all SACR individuals jointly.   
In Figure~\ref{fig:SACR_uncertainty}, the uncertainty in the predicted location for each individual is small when sufficient telemetry data exist, but increases as the sparsity of data increases resulting in the ``bumps'' in the uncertainty lines.  When the individuals lacking telemetry data are well connected to other individuals with more regular data, the potential for a reduction in uncertainty is greatest.  An example of uncertainty reduction occurred during the short period of time near September 16 when data existed for all SACR individuals except the orange individual (bottom plot in Figure~\ref{fig:SACR_uncertainty}).  In this case, we see a reduction in the uncertainty for the orange individual because it was well connected to at least one other individual at that time according to posterior individual network degree (Figure~\ref{fig:SACR_deg}).  Thus, a knowledge of all individuals in a migratory group helps account for changes in movement dynamics and can reduce uncertainty in the predicted locations of individuals.  By contrast, the purple individual became more disconnected from the other individuals throughout the migration and therefore Figure~\ref{fig:SACR_uncertainty} indicates no reduction of uncertainty when fitting the multiple individual model.   

\section{Conclusion}
Convolution specifications for continuous-space models have been popular in spatial statistics (Higdon, 2002; Ver Hoef and Peterson, 2010), but they have only recently been applied to model animal trajectories (e.g., Hooten and Johnson, 2017a; Scharf et al., 2018).  We demonstrated how convolution-based statistical models for trajectories can be useful to model the trajectories of migratory birds.  To account for heterogeneity in the dynamics of animal trajectories we introduced a flexible and mechanistically linked temporal warping function that can improve inference on individual trajectories as well as provide quantitative insight about the timing and duration of migration periods.  The process convolution approach to movement modeling could also be useful for identifying migration corridors using telemetry data from multiple individuals (e.g., Sawyer et al., 2009; Buderman et al., 2016).

Following the convolution nesting approach of Scharf et al. (2018), we used three stages of convolutions to account for time-varying dynamics in individual trajectories (without relying on the temporal warping approach described previously for individual trajectories).  The nesting of convolutions is particularly useful for characterizing the migratory behavior of groups of birds because they may change their social network structure during different portions of the migration (e.g., the clustering of SACR individuals we observed during stopovers).  Furthermore, the migratory network movement model may improve the inference pertaining to geographic position because it leverages the potential connectivity to borrow strength from individuals with more, or higher quality, data to assist the inference for individuals with missing data.  This concept could be used to design optimal duty cycling for telemetry devices for groups of moving individuals to save battery power and extend the life of the device providing more data for movement ecology studies.  Our approach to account for dependence among individuals in movement models is a model-based analog to the concept of ``cokriging,'' where statistical prediction of multivariate quantities is of interest (Ver Hoef and Barry, 1998).  Thus, similar methods can be used to model other multivariate spatio-temporal phenomena like atmospheric and geological processes.  

Our methods rely on well-known Gaussian process specifications and we leveraged common techniques in big data settings to implement the models and improve inference.  The temporal deformation approach we described has ties to spatial statistics (Sampson and Guttorp, 1992; Schmidt and O'Hagan, 2003) and provides an accessible way to fit nonstationary Gaussian process models using Bayesian model averaging.  To compute posterior model probabilities, we applied the two-stage procedure developed by Barker and Link (2013) that allowed us to fit individual movement models and then post-process model output to compare individual movement models in our GWFG example.         

Overall, while discrete-time animal movement models are still commonly employed, continuous-time continuous-space models for animal movement are useful when data are collected irregularly in time and continuous-time inference is desired.  By extending continuous-time movement models to accommodate heterogeneous dynamics, we showed that convolution specifications provide a valuable means to characterize complex trajectories of migratory animals.    

\section*{Acknowledgements}
The authors thank the Methods in Ecology and Evolution editors and reviewers for several suggestions that helped improved this manuscript and A. Walsh, B. Ballard, and J. VonBank for assistance deploying the tracking devices on Greenland white-fronted geese in Ireland.
This research was funded by NSF DMS 1614392.  Any use of trade, firm, or product names is for descriptive purposes only and does not imply endorsement by the U.S. Government.

\section*{Author Contributions}
All authors contributed to the ideas and methodology; MW and AP collected the data; MH and HS analysed the data; MH, HS, and TH led the writing of the manuscript. All authors contributed critically to the drafts and gave final approval for publication.

\section*{References}

\rf Auger-M\'eth\'e, M., A.E. Derocher, C.A. DeMars, M.J. Plank, E.A. Codling, and M.A. Lewis. (2016). Evaluating random search strategies in three mammals from distinct feeding guilds.  Journal of Animal Ecology, 85: 1411-1421.

\rf Banerjee, S., A.E. Gelfand, A.O. Finley, and H. Sang.  (2008).  Gaussian predictive process models for large spatial data sets.  Journal of the Royal Statistical Society: Series B, 70: 825-848. 

\rf Barker, R., and W. Link. (2013), Bayesian multimodel inference by RJMCMC: A Gibbs sampling approach. The American Statistician, 67: 150-156.

\rf Barry, R.P., and J. Ver Hoef. (1996). Blackbox kriging: spatial prediction without specifying variogram models. Journal of Agricultural, Biological, and Environmental Statistics 1: 297-322.

\rf Bauer, S. and M. Klaassen. (2013). Mechanistic models of animal migration behaviour-their diversity, structure and use. Journal of Animal Ecology, 82: 498-508.

\rf Berliner, L. (1996), Hierarchical Bayesian Time Series Models, in Maximum Entropy and Bayesian Methods, eds. K. Hanson, and R. Silver, Dordrecht, The Netherlands: Kluwer Academic Publishers, pp. 15-22.

\rf Buderman, F.E., M.B. Hooten, J.S. Ivan, and T.M. Shenk. (2016). A functional model for characterizing long distance movement behavior. Methods in Ecology and Evolution, 7: 264-273. 

\rf Brillinger, D.R. (2010). Modeling spatial trajectories. In Gelfand, A. E., P. J. Diggle, M. Fuentes, and P. Guttorp, editors, Handbook of Spatial Statistics, chapter 26, pages 463—-475. Chapman \& Hall/CRC, Boca Raton, Florida, USA.  

\rf Cagnacci, F., L. Boitani, R.A. Powell, and M.S. Boyce.  (2010).  Animal ecology meetings GPS-based radiotelemetry:  A perfect storm of opportunities and challenges.  Philosophical Transactions of the Royal Society of London B: Biological Sciences, 365: 2157-2162.

\rf Cagnacci, F., S. Focardi, A. Ghisla, B. Moorter, E.H. Merrill, E. Gurarie, M. Heurich, A. Mysterud, J. Linnell, M. Panzacchi, and R. May. (2016).  How many routes lead to migration?  Comparison of methods to assess and characterize migratory movements. Journal of Animal Ecology, 85: 54-68.

\rf Cangelosi, A.R. and M.B. Hooten (2009). Models for Bounded Systems with Continuous Dynamics. Biometrics, 65: 850-856.

\rf Cooke, F., D.B. Lank, and R.F. Rockwell. (1995). The Snow Geese of La Perouse Bay: Natural Selection in the Wild. Oxford University Press, Oxford, UK.

\rf Comiso, J.C. and C.L. Parkinson. (2004). Satellite-observed changes in the Arctic. Physics Today, 57: 38-44.

\rf Dalgety, C.T. and P. Scott. (1948).  A new race of the white-fronted goose.  Bulletin of the British Ornithologists’ Club, 68: 109-121.

\rf Datta, A., S. Banerjee, A.O. Finley, and A.E. Gelfand. (2016). Hierarchical nearest-neighbor Gaussian process models for large geostatistical datasets. Journal of the American Statistical Association, 111: 800-812. 

\rf Dunn, J. and P. Gipson. (1977).  Analysis of radio-telemetry data in studies of home range. Biometrics, 33: 85-101.

\rf Easterling, M.R., S.P. Ellner, and P.M. Dixon. (2000). Size-specific sensitivity: applying a new structured population model.  Ecology, 81: 694-708.

\rf Ellner, S.P. and M. Rees. (2006). Integral projection models for species with complex demography. The American Naturalist, 167: 410-428.

\rf Ely, C.R., A.D. Fox, R.T. Alisauskas, A. Andreev, R.G. Bromley, A.G. Degtyarev, B. Ebbinge, E.N. Gurtovaya, R. Kerbes, Alexander V. Kondratyev, I. Kostin, A.V. Krechmar, K.E. Litvin, Y. Miyabayashi, J.H. Moou, R.M. Oates, D.L. Orthmeyer, Y. Sabano, S.G. Simpson, D.V. Solovieva, Michael A. Spindler, Y.V. Syroechkovsky, J.Y. Takekawa, and A. Walsh. (2005). Circumpolar variation in morphological characteristics of greater white-fronted geese \emph{Anser albifrons}. Bird Study, 52: 104-119.

\rf Farine, D.R. (In Press).  When to choose dynamic vs. static social network analysis.  Journal of Animal Ecology.

\rf Farine, D.R. and H. Whitehead. (2015). Constructing, conducting and interpreting animal social network analysis. Journal of Animal Ecology, 84: 1144-1163.

\rf Fleming, C.H., J.M. Calabrese, T. Mueller, K.A. Olson, P. Leimgruber, and W.F. Fagan. (2014). Non-Markovian maximum likelihood estimation of autocorrelated movement processes.  Methods in Ecology and Evolution, 5: 462-472.

\rf Fox, A.D. and D.A. Stroud. (1988). The breeding biology of the Greenland white-fronted goose (\emph{Anser albifrons flavirostris}). Meddelelser Om Gronland, Bioscience, 27: 1-14.

\rf Fox, A.D. and D.A. Stroud. (2002). Greenland white-fronted goose \emph{Anser albifrons flavirostris}. Birds of the Western Palearctic Update, 4: 65-88.

\rf Fox, A.D., J.O. Hilmarsson, O. Einarsson, H. Boyd, J.N. Kristiansen, D.A. Stroud, A.J. Walsh, S.M. Warren, C. Mitchell, I.S. Francis, T. Nygard. (1999). Phenology and distribution of Greenland white-fronted geese \emph{Anser albifrons flavirostris} staging in Iceland. Wildfowl, 50: 29-43.

\rf Fox, A.D., I. Francis, D. Norriss, and A. Walsh. (2016). Report of the 2015/16 International census of Greenland white-fronted geese. Greenland White-fronted Goose Study, Ronde, Denmark.

\rf Fox, A.D. and A.J. Walsh.  (2012).  Warming winter effects, fat store accumulation and timing of spring departure of Greenland white-fronted geese \emph{Anser albifrons flavirostris} from their winter quarters.  Hydrobiologia, 697: 97-102.

\rf Fox, A.D., M.D. Weegman, S. Bearhop, G.M. Hilton, L. Griffin, D.A. Stroud, and A. Walsh. (2014). Climate change and contrasting plasticity in timing of a two-step migration episode of an Arctic-nesting avian herbivore. Current Zoology, 60: 233-242.

\rf Furrer, R., M.G. Genton, and D. Nychka.  2006. Covariance tapering for interpolation of large spatial datasets. Journal of Computational and Graphical Statistics, 15: 502-523.

\rf Goldenberg, A., A.X. Zheng, S.E. Fienberg, and E.M. Airoldi. (2010). A survey of statistical network models. Foundations and Trends in Machine Learning, 2: 129-233.

\rf Gurarie, E., F. Cagnacci, W. Peters, C.W. Fleming, J.M. Calabrese,T. Mueller, and W.F. Fagan. (2017a). A framework for modelling range shifts and migrations: Asking when, whither, whether and will it return. Journal of Animal Ecology, 86: 943-959.

\rf Gurarie, E., C.H. Fleming, W.F. Fagan, K.L. Laidre, J. Hernandez-Pliego, and O. Ovaskainen. (2017b). Correlated velocity models as a fundamental unit of animal movement: Synthesis and applications. Movement Ecology, 5: 13.

\rf Gurarie, E. and O. Ovaskainen. (2011). Characteristic spatial and temporal scales unify models of animal movement. The American Naturalist, 178: 113-123.

\rf Hanks, E.M., M.B. Hooten, and M. Alldredge. (2015). Continuous-time discrete-space models for animal movement. Annals of Applied Statistics, 9: 145-165. 

\rf Hays, G.C., L.C. Ferreira, A.M. Sequeira, M.G. Meekan, C.M. Duarte, H. Bailey, F. Bailleul, W.D. Bowen, M.J. Caley, D.P. Costa, V.M. Eguiluz, et al. (2016). Key questions in marine megafauna movement ecology. Trends in Ecology and Evolution, 31: 463-475.

\rf Hefley, T.J., K.M. Broms, B.M. Brost, F.E. Buderman, S.L. Kay, H.R. Scharf, J.R. Tipton, P.J. Williams, and M.B. Hooten. (2017). The basis function approach to modeling autocorrelation in ecological data. Ecology, 98: 632-646.

\rf Higdon, D. (2002). Space and space-time modeling using process convolutions. Pages 38–56 in C. W. Anderson, V. Barnett, P. C. Chatwin, and A. H. El-Shaarawi, editors. Quantitative methods for current environmental issues. Springer, New York, New York, USA.

\rf Hobbs, N.T. and M.B. Hooten. (2015). Bayesian Models: A Statistical Primer for Ecologists. Princeton University Press.

\rf Hoff, P.D., A.E. Raftery, and M.S. Handcock. (2002). Latent space approaches to social network analysis. Journal of the American Statistical Association, 97: 1090-1098.

\rf Hooten, M.B., Johnson, D.S., Hanks, E.M., and J.H. Lowry. (2010). Agent-based inference for animal movement and selection. Journal of Agricultural, Biological and Environmental Statistics, 15: 523-538.

\rf Hooten, M.B., F.E. Buderman, B.M. Brost, E.M. Hanks, and J.S. Ivan. (2016). Hierarchical animal movement models for population-level inference. Environmetrics, 27: 322-333. 

\rf Hooten, M.B. and D.S. Johnson. (2017a). Basis function models for animal movement. Journal of the American Statistical Association, 112: 578-589.

\rf Hooten, M.B. and D.S. Johnson. (2017b). Modeling Animal Movement. Gelfand, A.E., M. Fuentes, and J.A. Hoeting (eds). In Handbook of Environmental and Ecological Statistics. Chapman and Hall/CRC. 

\rf Hooten, M.B., D.S. Johnson, B.T. McClintock, and J.M. Morales. (2017). Animal Movement: Statistical Models for Telemetry Data. Chapman and Hall/CRC.

\rf Hooten, M.B. and N.T. Hobbs. (2015). A guide to Bayesian model selection for ecologists. Ecological Monographs, 85: 3-28.

\rf Jacoby, D.M. and R. Freeman. (2016). Emerging network-based tools in movement ecology. Trends in Ecology and Evolution, 31: 301-314.

\rf Johnson, D.S., J.M. London, M.A. Lea, and J.W. Durban. (2008). Continuous-time correlated random walk model for animal telemetry data. Ecology, 89: 1208-1215.

\rf Kays, R., M. Crofoot, W. Jetz, and M. Wikelski.  (2015).  Terrestrial animal tracking as an eye on life and planet.  Science, 384(6240): aaa2478. 

\rf Krapu, G. L., D. A. Brandt, K. L. Jones, and D. H. Johnson. 2011. Geographic distribution of the mid-continent population of sandhill cranes and related management applications. Wildlife Monographs, 175: 1-38.

\rf Krapu, G. L., D. A. Brandt, P. J. Kinzel, and A. T. Pearse. 2014. Spring migration ecology in the mid-continent population of sandhill cranes with an emphasis on the central Platte River Valley, Nebraska. Wildlife Monographs, 189: 1-44.

\rf Kruse, K. L., and J. A. Dubovsky. 2015. Status and harvests of sandhill cranes: Mid-continent, Rocky Mountain, Lower Colorado River Valley and Eastern populations. U.S. Fish and Wildlife Service, Lakewood, Colorado.

\rf Langrock, R., R. King, J. Matthiopoulos, L. Thomas, D. Fortin, and J. Morales.  (2012).  Flexible and practical modeling of animal telemetry data: Hidden Markov models and extensions.  Ecology, 93: 2336-2342.

\rf Madsen, J. and A.D. Fox. (1981). The summer diet of the Greenland White-fronted Goose. Report of the 1979 Expedition to Eqalungmiut Nunat, West Greenland (ed A.D. Fox and D.A. Stroud), pp. 108-118. Greenland White-fronted Goose Study, Aberystwyth, UK.

\rf Malecki, R.A., A.D. Fox, and B.D.J. Batt. (2000).  An aerial survey of nesting Greenland white-fronted and Canada geese in west Greenland. Wildfowl, 51: 49-58.

\rf McClintock, B.T., R. King, L. Thomas, J. Matthiopoulos, B.J. McConnell, and J.M. Morales. (2012). A general discrete‐time modeling framework for animal movement using multistate random walks. Ecological Monographs, 82: 335-349.

\rf McClintock, B.T., D.S. Johnson, M.B. Hooten, J.M. Ver Hoef, and J.M. Morales. (2014). When to be discrete: the importance of time formulation in understanding animal movement. Movement Ecology, 2: 21. 

\rf McKellar, A.E., R. Langrock, J.R. Walters, and D.C. Kesler. 2014. Using mixed hidden Markov models to examine behavioral states in a cooperatively breeding bird. Behavioral Ecology, 26: 148-157.

\rf Morales, J.M., D.T. Haydon, J. Frair, K.E. Holsinger, and J.M. Fryxell.  (2004). Extracting more out of relocation data:  Building movement models as mixtures of random walks. Ecology, 85: 2436-2445.

\rf Parton, A. and P.G. Blackwell. (2017). Bayesian inference for multistate `step and turn' animal movement in continuous time.  Journal of Agricultural, Biological, and Environmental Statistics, 22: 373-392.

\rf Pearse, A.T., G.L. Krapu, D.A. Brandt, and G.A. Sargeant. 2015. Timing of spring surveys for midcontinent sandhill cranes. Wildlife Society Bulletin, 39: 87-93.

\rf Peterson, E.E. and J.M. Ver Hoef. (2010). A mixed-model moving‐average approach to geostatistical modeling in stream networks. Ecology, 91: 644-651.

\rf Postlethwaite, C.M., P. Brown, and T.E. Dennis.  2013.  A new multi-scale measure for analysing animal movement data.  Journal of Theoretical Biology, 317: 175-185.  

\rf Russell, J.C., E.M. Hanks, and D. Hughes.  (2017).  Modeling collective animal movement through interactions in behavioral states.  Journal of Agricultural, Biological, and Environmental Statistics, In Press.

\rf Russell, J.C., E.M. Hanks, and M. Haran. (2017). Dynamic models of animal movement with spatial point process interactions. Journal of Agricultural, Biological, and Environmental Statistics, 21: 22-40.

\rf Ruttledge, R.F. and M.A. Ogilvie. (1979). The past and current status of the Greenland white-fronted goose in Ireland and Britain. Irish Birds, 1: 293-363.

\rf Sampson, P. (2010). Constructions for nonstationary spatial processes. Gelfand, A.E., P.J. Diggle, M. Fuentes, and P. Guttorp (eds). In Handbook of Spatial Statistics. Chapman and Hall/CRC. 

\rf Sampson, P. and P. Guttorp. (1992).  Nonparametric estimation of nonstationary spatial covariance structure. Journal of the American Statistical Association, 87: 108-119.

\rf Sawyer, H., M.J. Kauffman, R.M. Nielson, and J.S. Horne. (2009). Identifying and prioritizing ungulate migration routes for landscape-level conservation. Ecological Applications, 19: 2016-2025.

\rf Scharf, H.R., M.B. Hooten, B.K. Fosdick, D.S. Johnson, J.M. London, and J.W. Durban. (2016). Dynamic social networks based on movement. Annals of Applied Statistics, 10: 2182-2202.

\rf Scharf, H.R., M.B. Hooten, D.S. Johnson, and J.W. Durban. (2017). Imputation approaches for animal movement modeling. Journal of Agricultural, Biological, and Environmental Statistics, 22: 335-352.

\rf Scharf, H.R., M.B. Hooten, D.S. Johnson, and J.W. Durban. (2018).  Process convolution approaches for modeling interacting trajectories.  Environmetrics, e2487.

\rf Schmidt, A. and A. O'Hagan. (2003). Bayesian inference for nonstationary spatial covariance structure via spatial deformations. Journal of the Royal Statistical Society, Series B, 65: 743-758.

\rf Soleymani, A., F. Pennekamp, S. Dodge, and R. Weibel. (2017). Characterizing change points and continuous transitions in movement behaviours using wavelet decomposition.  Methods in Ecology and Evolution, 8: 1113-1123.

\rf Stoll, J. R., R. B. Ditton, and T. L. Eubanks. 2006. Platte River birding and the spring migration: Humans, values, and unique ecological resources. Human Dimensions of Wildlife 11:241-254.

\rf Stroud, D.A., A.D. Fox, C. Urquhart and I.S. Francis. (compilers) (2012). International Single Species Action Plan for the Conservation of the Greenland White-fronted Goose \emph{Anser albifrons flavirostris}, 2012-2022.  AEWA Technical Series No. 45, Bonn, Germany.

\rf Ver Hoef, J.M., and R. Barry.  (1998).  Constructing and fitting models for cokriging and multivariable spatial prediction.  Journal of Statistical Planning and Inference, 69: 275-294.

\rf Ver Hoef, J.M., and E. Peterson.  (2010).  A moving average approach for spatial statistical models of stream networks.  Journal of the American Statistical Association, 105: 6-18.

\rf Weegman, M.D., A.D. Fox, G.M. Hilton, D.J. Hodgson, A.J. Walsh, L.R. Griffin, and S. Bearhop. (2017). Diagnosing the decline of the Greenland white-fronted goose Anser albifrons flavirostris using population and individual level techniques. Wildfowl, 67: 3-18.

\rf Wheeler, R.H., and J.C. Lewis. 1972. Trapping techniques for sandhill crane studies in the Platte River Valley.  U.S. Department of Interior, U.S. Fish and Wildlife Service Publication 107.  Washington D.C., USA.

\rf Whoriskey, K., M. Auger-M\'eth\'e, C.M. Albertsen, F.G. Whoriskey, T.R. Binder, C.C. Krueger, and J. Mills Flemming. (2017). A hidden Markov movement model for rapidly identifying behavioral states from animal tracks. Ecology and Evolution, 7: 2112-2121.

\rf Wikle, C. (2010).  Low-Rank Representations for Spatial Processes, in Handbook of Spatial Statistics, eds. A. Gelfand, P. Diggle, M. Fuentes, and P. Guttorp, Boca Raton, FL: Chapman \& Hall/CRC, pp. 107-118. 

\pagebreak

\pagebreak

\pagebreak

\pagebreak

\pagebreak

\pagebreak

\pagebreak

\pagebreak
\section*{Appendix S1:  Glossary}
\begin{sidewaystable}[htp]
{
\footnotesize
\begin{tabular}{ll}
\hline 
Term & Definition\tabularnewline
\hline 
  Convolution & An integral of the product of two functions with respect to the same variable  \\
  Gaussian Process & A multivariate Normal random variable with dependence expressed using covariance \\
  Kernel function & A positive function providing weights for averaging another process (e.g., a normal distribution) \\
  Measurement error & Location uncertainty associated with telemetry data that are not equal to the true positions \\
  Movement trajectory & The track associated with an individual animal's movement (continuous set of positions over time) \\
  Nested process convolution & Sequentially convolving a function with another convolution from the same or different stochastic process \\
  Nonstationary & Heterogeneity of the statistical properties associated with a process change over time (or space)  \\
  Predicted trajectory & A statistical estimate of the true movement trajectory (and possibly its probability distribution) \\
  Process convolution & A covolution of a kernel function and a stochastic process, usually serving as a model for a latent process \\
  Stochastic process & An infinite number of random variables in a domain (e.g., time), possibly correlated \\
  Telemetry data & Recorded locations of animal positions over time  \\
  Temporal deformation & A function that warps (expands or compresses) time \\
\hline 
\end{tabular}
}
\end{sidewaystable}

\pagebreak
\section*{Appendix S2:  GWFG Model Implementation}

\begin{enumerate}
  \item Project Data: We converted the telemetry data $\mathbf{s}(t_i)$ ($i=1,\ldots,n$) to an equidistant projection centered on the geographical mean of the data.  Center and scale projected telemetry data by subtracting mean and dividing by pooled geographic standard deviation.  Scale time domain so that $t_1=0$ and $t_n=1$; this scaling makes it easy to specify reasonable priors for parameters controlling the range of $g(t,\tau)$.    
  \item Specify Convolution Type: We calculated the kernel functions at a large, but finite, number of $m$ time points to closely approximate the trajectory process.  In the GWFG example, we used $m=800$ and a Gaussian probability density function for $g(t,\tau)$ with scale parameter $\phi$.  
  \item Specify Likelihood:  We specified the integrated model for the data as in Eqn. 3 based on the matrices $\mathbf{H}$ resulting from the choice of $g(t,\tau)$:   
  \begin{equation}
    \mathbf{s} \sim \text{N}(\boldsymbol\mu_0\otimes \mathbf{1}, \sigma^2_s (\mathbf{I} + \sigma^2_{\mu/s} (\mathbf{I}\otimes \mathbf{H})(\mathbf{I}\otimes \mathbf{H})'))\;, 
  \end{equation}
  where $\sigma^2_s$ corresponds to the measurement error variance and $\sigma^2_{\mu/s}\equiv\sigma^2_\mu/\sigma^2_s$ with $\sigma^2_\mu$ representing the process variance.  We set $\boldsymbol\mu_0 \equiv\mathbf{s}(t_1)$.  
  \item Specify Priors: We used the following priors for $\sigma^2_s\sim \text{Inverse Gamma}(1.0558\times10^{-10},2)$, $\sigma_{\mu/s}\sim \text{Unif}(0,20)$, and $\phi\sim \text{DiscUnif}(\Phi)$, where $\Phi$ are $100$ equally spaced real numbers between $0.001$ and $0.02$.  The prior for the measurement error variance was informative to represent the well-known GPS accuracy to within 10 meters of the true position (converted to our rescaled spatial domain for analysis).  The discrete uniform prior for $\phi$ allowed us to pre-calculate the matrix $(\mathbf{I}\otimes \mathbf{H})(\mathbf{I}\otimes \mathbf{H})'$ and recall as necessary in the MCMC algorithm.   
  \item Implement Temporal Deformation:  To account for heterogeneity in the movement dynamics, we implemented the mixture model using separate model fits based on $f_l(t)$ equally spaced truncated Gaussian kernels $f(t)$ throughout the time domain.  We also explored the space of the kernel scale parameter $\phi_l$ as well as the warp magnitude parameter $\sigma^2_w$ using a grid search over 100 permutations of values for these parameters (ranging from $0.01$--$0.0625$ for $\phi_l$ and $0.6$--$0.8$ for $\sigma^2_w$).  Based on a preliminary model fit without temporal deformation, we scored the set of $100\times 100$ warping functions using the deviance score (Hooten and Hobbs, 2015) and fit the full model to the resulting top 20 using an MCMC algorithm.  We used second-stage Bayesian model averaging (Barker and Link, 2013) to find the posterior model probabilities.  
  \item Prediction: We sampled realizations from the model averaged posterior predictive trajectory distribution to create Figure 4.  
\end{enumerate}

Using sequential programming, the total computational time required to fit the model to the telemetry data for an individual, compute the posterior model probabilities, and obtain predictions was approximately 45 minutes on a computer with 3Ghz processors and 64 GB of RAM.  Because of the Gaussian process specification for the model with a covariance matrix that can be inverted using the Sherman-Morrison-Woodbury method, the required computing time would not increase substantially for larger data sets.  Using parallel programming, the computing time could be reduced further depending on how many processors were used.  

Bayesian model averaged posterior means and standard deviations for model parameters for each individual (correponding to colors in the GWFG figures) based on scaled data are shown in the table below:

\begin{table}[htp]
  \centering
  \footnotesize
  \begin{tabular}{c c c c c c c c c}
    \multicolumn{1}{c}{} & \multicolumn{2}{c}{red} & \multicolumn{2}{c}{blue} & \multicolumn{2}{c}{green} & \multicolumn{2}{c}{purple} \\
    \hline
    parameter    & mean & sd & mean & sd &  mean & sd & mean & sd \\
    \hline
     $\sigma^2_s$  & 0.000032 & 0.000078 & 0.000018 & 0.0000046 & 0.000001 & 0.0000097 & 0.000016 & 0.0000033    \\
     $\sigma^2_\mu$  & 0.0034 & 0.0011 & 0.0065 & 0.0008 & 0.0032 & 0.0003 & 0.0056 & 0.0008      \\
     $\phi$  & 0.003 & 0.0017 & 0.019 & 0.0008 & 0.007 & 0.0003 & 0.019 & 0.0011     \\
     \hline
  \end{tabular}
  \label{tab:results}
\end{table}

\pagebreak
\section*{Appendix S3:  SACR Model Implementation}
\begin{enumerate}
  \item Project Data: We converted the telemetry data $\mathbf{s}(t_i)$ $(i=1,\ldots,n)$ to an equidistant projection centered on -110$^{\circ}$ longitude, 51$^{\circ}$ latitude. We scaled the time domain so that $t_1=0$ and $t_n=1$; this scaling makes it easy to specify reasonable priors for parameters controlling the temporal range of the kernel functions $h^{(2)}(t,\tau)$ and $h_z(t,\tau)$.
  \item Specify Nested Convolution Types: We evaluated the process convolution over a dense, but finite, grid of $m=260$ equally-spaced time points to closely approximate the trajectory process. This grid corresponds to approximately 5 times points per day. In the SACR example, we used a Gaussian probability density function for for $h^{(2)}_{jk}(t, \tau)$ with scale parameter $\phi$. We used the kernel given in (Eqn. 10), conditioned on the unobserved migratory network weights $\boldsymbol\nu$, to induce dependence across individuals.  
  \item Specify Latent Migratory Network Model:  To account for a network that smoothly varies over time, we specified a prior for the latent processes $\mathbf{z}_j(t)$ (which act as random effects in the model).  In the latent space network model, $\mathbf{z}_j(t)$ is a point that we assume exists on a latent 2 dimensional real plane (${\cal Z}$, where the units pertain to the latent space only and not the geographic space).  Thus, because the set $\mathbf{z}_j(t)$ for all times $t$ forms a latent trajectory, we used a convolution-based movement model as a prior for the latent point processes themselves in the latent space.  In essence, this prior is a movement model within a movement model.  Specifying the network weights in terms of a set of latent points with a prior is simply a mechanism to reduce the dimensionality of the network to something that can be estimated.  For the SACR migration network example, we specified
\begin{equation}
  \mathbf{z}_{j}(t) = \int_{t_1}^{t_n} \sigma_z \mathbf{H}_{z}(t,\tau)d\mathbf{b}_j(\tau) \;, \label{eq:z} \\
\end{equation}
\noindent where $\mathbf{H}_{z}(t,\tau)$ has diagonal elements $h_z(t,\tau)= \exp(-(t-\tau)^2/\phi_z)$.  This stochastic process serves as a prior for the network, where the hyperparameter $\sigma_z$ controls the overall density of the network and $\phi_z$ controls the rate of change in the dynamics of the network.  Larger values of $\sigma_z$ imply a less connected network over time because the $\sigma_z$ pushes the $\mathbf{z}_{j}(t)$ away from each other in latent space, whereas, large values of $\phi_z$ imply a more slowly varying network in time.  The stochastic process $d\mathbf{b}_j(\tau)$ is assumed to be white noise with variance $d\tau$.  The hyperparameters $\sigma_z$ and $\phi_z$ represent our understanding of the variation in network structure over time (so that network connections change less than 10 times per study period on average yet also allows for a realistic range of network densities).  We approximated the process convolution over a grid of $m_w=15$ time points, corresponding to approximately 2 time points per week. The sparsity of the time grid for the latent migratory network process relative to the observed movement process is appropriate because the latent process evolves much more slowly in time and does not required as dense a grid to adequately approximate the necessary integral.  We used a Gaussian function for $h_z(t,\tau)$ with scale hyperparameter $\phi_z=0.08$ and network density hyperparameter $\sigma_z=10$.
  \item Specify Likelihood: We specified the integrated model for the data as in Eqn~2 based on the matrices $\mathbf{H}^{(1)}$, $\mathbf{H}^{(2)}$, and $\mathbf{H}^{(3)}$ resulting from the choice of $h^{(1)}_{jk}(t,\tau)$, $h^{(2)}_{jk}(t,\tau)$, and $h^{(3)}_{jk}(t,\tau)$. The $2J\times 1$ vector $\boldsymbol{\mu}_0$ contains the initial locations for all $J=5$ individuals. The resulting likelihood is:
\[
  \mathbf{s} \sim \text{N} \left( 
    \boldsymbol{\mu}_0 \otimes \mathbf{1}_m, 
    \sigma_s^2 \left(
      \mathbf{I}_{2mJ} + \sigma_{\mu/s}^2 \left( \mathbf{I}_2 \otimes
        \mathbf{H}^{(3)} \mathbf{H}^{(2)} \mathbf{H}^{(1)} \right)
      \left(\mathbf{I}_2 \otimes
        \mathbf{H}^{(3)} \mathbf{H}^{(2)} \mathbf{H}^{(1)} \right)'
    \right)
  \right)
\]
where $\sigma_s^2$ corresponds to the measurement error variance and $\sigma_{\mu/s}^2 \equiv \sigma_\mu^2 / \sigma_s^2$ with $\sigma_\mu^2$ representing the process variance. 
  \item Specify Priors: We used the following priors for $\sigma_s^2 \sim \text{Inverse Gamma}(10^{-3}, 10^{-3})$,  $\sigma_{\mu/s}^2 \sim \text{Inverse Gamma}(10^{-3}, 10^{-3})$, $\phi \sim \text{Gamma}(2, 200)$. Additionally, we used the following priors for $\boldsymbol{\mu}_0 \sim \text{N}(\mathbf{0}, \sigma_0^2\mathbf{I})$, $\sigma_0^2 \sim \text{Inverse Gamma}(1, 10)$, and $\sigma_w^2 \sim \text{Inverse Gamma}(52, 10)$.
  \item Prediction: We sampled realizations from the posterior predictive trajectory distribution to create Figure 5.
\end{enumerate}

\end{document}